\documentclass[twocolumn,twoside]{IEEEtran}

\usepackage{cite}

\usepackage[applemac]{inputenc}
\usepackage{fontenc}
\usepackage{caption}
\usepackage{subcaption}
\usepackage{placeins}
\ifCLASSINFOpdf
  \usepackage[pdftex]{graphicx}
\else
   \usepackage[dvips]{graphicx}
\fi

\usepackage{graphicx}
\usepackage{caption}
\usepackage{subcaption}

\usepackage[cmex10]{amsmath}
\usepackage{url}
\usepackage{pdfsync}
\usepackage{dsfont}
\usepackage{float}
\usepackage{amssymb}
\usepackage{acronym,xcolor}
\usepackage[squaren,Gray]{SIunits}
\usepackage{amsmath}
\usepackage{comment}

\acrodef{$P_{EM}$}{probability of emulation, or false alarm}
\acrodef{$P_{FA}$}{probability of false alarm}
\acrodef{$P_{MD}$}{probability of missed detection}
\acrodef{$P_{D}$}{probability of detection}

\acrodef{ACF}{autocorrelation function}
\acrodef{ACG}{automatic	gain control}
\acrodef{ACI}{adjacent channel interference}
\acrodef{ACK}{acknowledge}
\acrodef{AcR}{autocorrelation receiver}
\acrodef{ADC}{analog-to-digital converter}
\acrodef{AF}{amplify \& forward}
\acrodef{AFL}{anchor-free localization}
\acrodef{AGNSS}{assisted-GNSS}
\acrodef{AGPS}{assisted GPS}
\acrodef{AI}{automatic identification}
\acrodef{AIC}{Akaike information criterion}
\acrodef{AoA}{angle-of-arrival}
\acrodef{AoD}{angle-of-departure}
\acrodef{AOT}{approximate optimum threshold}
\acrodef{AP}{access point}
\acrodef{API}{application programming interface}
\acrodef{ASK}{amplitude shift keying}
\acrodef{ASNR}{accumulated signal-to-noise ratio}
\acrodef{AUB}{asymptotic union bound}
\acrodef{AWGN}{additive white Gaussian noise}
\acrodef{BAN}{body area network}
\acrodef{BAV}{balanced antipodal Vivaldi}
\acrodef{BCH}{Bose Chaudhuri Hocquenghem}
\acrodef{BEP}{bit error probability}
\acrodef{BER}{bit error rate}
\acrodef{BF}{brute force}
\acrodef{BFC}{block fading channel}
\acrodef{BIC}{Bayesian information criterion}
\acrodef{BLUE}{best linear unbiased estimator}
\acrodef{BPAM}{binary pulse amplitude modulation}
\acrodef{BPF}{bandpass filter}
\acrodef{BPPM}{binary pulse position modulation}
\acrodef{bps}{bits per second}
\acrodef{BPSK}{binary phase shift keying}
\acrodef{BPZF}{band-pass zonal filter}
\acrodef{BS}{base station}
\acrodef{BSC}{binary symmetric channel}
\acrodef{BTB}{Bellini-Tartara bound}
\acrodef{c.c.d.f.}{complementary cumulative distribution function}
\acrodef{c.d.f.}{cumulative distribution function}
\acrodef{CAD}{computer-aided design}
\acrodef{CAIC}{consistent Akaike information criterion}
\acrodef{CAP}{continuous aperture phased}
\acrodef{CCF}{cross correlation function}
\acrodef{CCI}{co-channel interference}
\acrodef{CD}{cooperative diversity}
\acrodef{CDMA}{code division multiple access}
\acrodef{CEOT}{channel ensemble optimum threshold}
\acrodef{CEP}{codeword error probability}
\acrodef{CFAR}{constant	 false alarm rate}
\acrodef{ch.f.}{characteristic function}
\acrodef{CH}{cluster head}
\acrodef{CIR}{channel impulse response}
\acrodef{CL}{centroid localization}
\acrodef{CM}{channel model}
\acrodef{CNR}{clutter-to-noise ratio}
\acrodef{CP}{ciclic prefix}
\acrodef{CPR}{channel pulse response}
\acrodef{CR}{channel response}
\acrodef{CRB}{Cram\'{e}r-Rao bound}
\acrodef{CRC}{cyclic redundancy check}
\acrodef{CRLB}{Cram\'{e}r-Rao lower bound}
\acrodef{CS}{clock skew}
\acrodef{CSCG}{circularly symmetric complex Gaussian}
\acrodef{CSI}{channel state information}
\acrodef{CSMA}{carrier sense multiple access}
\acrodef{CSS}{chirp spread spectrum}
\acrodef{CTS}{clear-to-send}
\acrodef{CW}{continuous wave}
\acrodef{DAA}{detect and avoid}
\acrodef{DAB}{digital audio broadcasting}
\acrodef{DBB}{digital base band}
\acrodef{DBPSK}{differential binary phase shift keying}
\acrodef{DCM}{dual-carrier modulation}
\acrodef{DDP}{detected direct path}
\acrodef{DF}{detect \& forward}
\acrodef{DFMS}{monopole dual feed stripline antenna}
\acrodef{DGPS}{differential GPS}
\acrodef{DLL}{delay-locked loop}
\acrodef{DoD}{Department of Defense}
\acrodef{DoF}{degrees of freedom}
\acrodef{DP}{direct path}
\acrodef{DR}{detection rate}
\acrodef{DRT}{distance ratio test}
\acrodef{DS-SS}{direct-sequence spread-spectrum}
\acrodef{DS}{delay spread}
\acrodef{DTR}{differential transmitted-reference}
\acrodef{DVB-H}{digital video broadcasting\,--\,handheld}
\acrodef{DVB-T}{digital video broadcasting\,--\,terrestrial}
\acrodef{e.m.}{electromagnetic}
\acrodef{ECC}{European Community Commission}
\acrodef{ED}{energy detector}
\acrodef{EDR}{energy detector receiver}
\acrodef{EFIM}{equivalent Fisher information matrix}
\acrodef{EIRP}{effective radiated isotropic power}
\acrodef{EKF}{extended Kalman filter}
\acrodef{ELP}{equivalent low-pass}
\acrodef{EM}{electromagnetic}
\acrodef{EMCB}{extended Miller Chang bound}
\acrodef{EME}{minimum eigenvalue ratio detector}
\acrodef{ENP}{estimated noise power}
\acrodef{ESA}{European Space Agency}
\acrodef{EU}{European Union}
\acrodef{FAR}{false alarm rate}
\acrodef{FCC}{Federal Communications Commission}
\acrodef{FDMA}{frequency division multiple access}
\acrodef{FDMA}{frequency division multiple access}
\acrodef{FEC}{forward error correction}
\acrodef{FEC}{forward error correction}
\acrodef{FFD}{full function device}
\acrodef{FFR}{full function reader}
\acrodef{FF}{far-field}
\acrodef{FFT}{fast Fourier transform}
\acrodef{FG}{factor graph}
\acrodef{FH-SS}{frequency-hopping spread-spectrum}
\acrodef{FH}{frequency-hopping}
\acrodef{FIM}{Fisher information matrix}
\acrodef{FLL}{Frequency-locked loop}
\acrodef{FS}{frame synchronization}
\acrodef{GA}{Gaussian approximation}
\acrodef{GD}{gradient descent}
\acrodef{GDOP}{geometric dilution of precision}
\acrodef{GLR}{generalized likelihood ratio}
\acrodef{GLRT}{generalized likelihood ratio test}
\acrodef{GML}{generalized maximum likelihood}
\acrodef{GPRS}{general packet radio service}
\acrodef{GPS}{global positioning system}
\acrodef{HAP}{high altitude platform}
\acrodef{HCRB}{hybrid Cram\'{e}r-Rao bound}
\acrodef{HDSA}{high-definition situation-aware}
\acrodef{Hi-RADIAL}{High-accuracy RAdio Detection, Identification, And Localization}
\acrodef{HMM}{hidden Markov model}
\acrodef{HPA}{high-power amplifier}
\acrodef{HPBW}{half power beam width}
\acrodef{HW}{hardware}
\acrodef{i.i.d.}{independent, identically distributed}
\acrodef{ICT}{information and communication technologies}
\acrodef{IE}{informative element}
\acrodef{IEEE}{Institute of Electrical and Electronics Engineers}
\acrodef{IF}{intermediate frequency}
\acrodef{IFFT}{inverse fast Fourier transform}
\acrodef{IMF}{ideal matched filter}
\acrodef{IMU}{inertial measurement unit}
\acrodef{INR}{interference-to-noise ratio}
\acrodef{INS}{inertial navigation system}
\acrodef{IoT}{Internet of things}
\acrodef{IIoT}{industrial Internet of things}
\acrodef{INS}{inertial navigation system}
\acrodef{IR-UWB}{impulse radio UWB}
\acrodef{IR}{impulse radio}
\acrodef{IRI}{inter-reader interference}
\acrodef{IRS}{intelligent reflecting surface} 
\acrodef{ISI}{inter-symbol interference} 
\acrodef{isi}{intra-symbol interference} 
\acrodef{ISM}{industrial, scientific and medical}
\acrodef{ISNR}{interference-plus-signal-to-noise-ratio}
\acrodef{IT}{interference temperature}
\acrodef{ITC}{information theoretic criteria}
\acrodef{JBSF}{jump back and search forward}
\acrodef{JF}{just forward}
\acrodef{KF}{Kalman filter}
\acrodef{LDC}{low duty cycle}
\acrodef{LDPC}{low density parity check}
\acrodef{LEO}{localization error outage}
\acrodef{LIS}{large intelligent surface}
\acrodef{LLR}{log-likelihood ratio}
\acrodef{LLRT}{log-likelihood ratio test}
\acrodef{LRT}{likelihood ratio test}
\acrodef{LNA}{low-noise amplifier}
\acrodef{LOS}{line-of-sight}
\acrodef{LRT}{likelihood ratio test}
\acrodef{LS}{least square}
\acrodef{LS}{least squares}
\acrodef{M-PSK}{$M$-ary phase shift keying}
\acrodef{M-QAM}{$M$-ary quadrature amplitude modulation}
\acrodef{m.g.f.}{moment generating function}
\acrodef{MAC}{medium access control}
\acrodef{MAE}{mean absolute error}
\acrodef{MAI}{multiple access interference}
\acrodef{MAN}{metropolitan area network}
\acrodef{MAP}{maximum a posteriori}
\acrodef{MB-OFDM}{multi-band OFDM}
\acrodef{MB-UWB}{multi-band UWB}
\acrodef{MB}{multi-band}
\acrodef{MC}{multi-carrier}
\acrodef{MCB}{Miller Chang bound}
\acrodef{MCRB}{modified Cram\'{e}r-Rao bound}
\acrodef{MDD}{minimum distance distribution}
\acrodef{MDL}{minimum description length}
\acrodef{MF}{matched filter}
\acrodef{MGF}{moment generating function}
\acrodef{MI}{mutual information}
\acrodef{MIMO}{multiple-input multiple-output}
\acrodef{MISO}{multiple-input single-output}
\acrodef{ML}{maximum likelihood}
\acrodef{MM}{min-max}
\acrodef{MME}{maximum-minimum eigenvalue ratio detector}
\acrodef{MMSE}{minimum mean-square error}
\acrodef{MPC}{multipath component}
\acrodef{MRC}{maximal ratio combiner}
\acrodef{MS}{mobile station}
\acrodef{MSB}{most significant bit}
\acrodef{MSE}{mean square error}
\acrodef{MSE}{mean squared error}
\acrodef{MSK}{minimum shift keying}
\acrodef{MUI}{multi-user interference}
\acrodef{MUR}{multistatic radar}
\acrodef{MVU}{minimum variance unbiased}
\acrodef{MZZB}{modified Ziv-Zakai bound}
\acrodef{NB}{narrowband}
\acrodef{NBI}{narrowband interference}
\acrodef{NEO}{navigation error outage}
\acrodef{NFER}{near-Þeld electromagnetic ranging}
\acrodef{NF}{near-field}
\acrodef{NFF}{near-field focused}
\acrodef{NL}{nonlinear}
\acrodef{NLOS}{non-line-of-sight}
\acrodef{NP}{Neyman-Pearson}
\acrodef{NTIA}{National Telecommunications and Information Administration}
\acrodef{NTP}{network time protocol}
\acrodef{OAM}{orbital angular momentum} 
\acrodef{OC}{optimum combining}
\acrodef{OFDM}{orthogonal frequency division multiplexing}
\acrodef{OOK}{on-off keying}
\acrodef{OP}{outage probability}
\acrodef{OT}{optimum threshold}
\acrodef{P-Max}{$P$-Max}  
\acrodef{p.d.f.}{probability density function}
\acrodef{p.m.f.}{probability mass function}
\acrodef{PA}{power amplifier}
\acrodef{PAM}{pulse amplitude modulation}
\acrodef{PAN}{personal area network}
\acrodef{PAR}{peak-to-average ratio}
\acrodef{PD}{probability of detection}
\acrodef{PDP}{power delay profile}
\acrodef{PE}{probability of emulation}
\acrodef{PEB}{position error bound}
\acrodef{PEP}{packet error probability}
\acrodef{PF}{particle filter}
\acrodef{PFA}{probability of false alarm}
\acrodef{PHY}{physical layer}
\acrodef{PL}{path-loss}
\acrodef{PLL}{phase-locked loop}
\acrodef{PMD}{probability of missed detection}
\acrodef{PN}{pseudo-noise}
\acrodef{ppm}{part-per-million}
\acrodef{PPM}{pulse position modulation}
\acrodef{PR}{pseudo-random}
\acrodef{PRake}{partial rake}
\acrodef{PRF}{pulse repetition frequency}
\acrodef{PRP}{pulse repetition period}
\acrodef{PSD}{power spectral density}
\acrodef{PSEP}{pairwise synchronization error probability}
\acrodef{PSK}{phase shift keying}
\acrodef{PSWF}{prolate spheroidal wave function}
\acrodef{PU}{primary user}
\acrodef{QAM}{quadrature amplitude modulation}
\acrodef{QoS}{quality of service}
\acrodef{QPSK}{quadrature phase shift keying}
\acrodef{R.V.}{random variable}
\acrodef{RADAR}{radar}
\acrodef{RCS}{radar cross section}
\acrodef{RDL}{"random data limit"}
\acrodef{REM}{radio environment map}
\acrodef{REO}{ranging error outage}
\acrodef{RF}{radio-frequency}
\acrodef{RFID}{radio-frequency identification}
\acrodef{RFR}{reduced function reader}
\acrodef{RFT}{reduced function tag}
\acrodef{RII}{ranging information intensity}
\acrodef{RIS}{reconfigurable intelligent surface}
\acrodef{rms}{root mean square}
\acrodef{RMSE}{root-mean-square error}
\acrodef{ROC}{receiver operating characteristic}
\acrodef{RRC}{root raised cosine}
\acrodef{RSN}{radar sensor network}
\acrodef{RSS}{received signal strength}
\acrodef{RSSI}{received signal strength indicator}
\acrodef{RTLS}{real time locating systems}
\acrodef{RTT}{round-trip time}
\acrodef{S-V}{Saleh-Valenzuela}
\acrodef{SA}{simulated annealing}
\acrodef{SaG}{stop-and-go}
\acrodef{SBS}{serial backward search}
\acrodef{SBSMC}{serial backward search for multiple clusters}
\acrodef{SCM}{supply chain management}
\acrodef{SCR}{signal-to-clutter ratio}
\acrodef{SEP}{symbol error probability}
\acrodef{SIS}{small intelligent surface}
\acrodef{SFD}{start frame delimiter}
\acrodef{SIMO}{single-input multiple-output}
\acrodef{SINR}{signal-to-interference plus noise ratio}
\acrodef{SIR}{signal-to-interference ratio}
\acrodef{SISO}{single-input single-output}
\acrodef{SNR}{signal-to-noise ratio}
\acrodef{SoC}{system on chip}
\acrodef{SoO}{signal of opportunity}
\acrodef{SoP}{system on package}
\acrodef{SOT}{sub-optimum threshold}
\acrodef{SPAWN}{sum-product algorithm over a wireless network}
\acrodef{SPEB}{squared position error bound}
\acrodef{SPMF}{single-path matched filter}
\acrodef{SQNR}{signal-to-quantization-noise ratio}
\acrodef{SS}{spread spectrum}
\acrodef{ST}{simple thresholding}
\acrodef{SU}{secondary user}
\acrodef{SVD}{singular value decomposition}
\acrodef{SW}{software}
\acrodef{SW}{sync word}
\acrodef{TDE}{time delay estimation}
\acrodef{TDL}{tapped delay line}
\acrodef{TDMA}{time division multiple access}
\acrodef{TDOA}{time difference-of-arrival}
\acrodef{TH}{time-hopping}
\acrodef{TNR}{threshold-to-noise ratio}
\acrodef{TOA}{Time-of-arrival}
\acrodef{TOF}{time-of-flight}
\acrodef{TPC}{transmit power control}
\acrodef{TR}{transmitted-reference}
\acrodef{TS}{tabu search}
\acrodef{UAV}{unmanned aerial vehicle}
\acrodef{UB}{union bound}
\acrodef{UDP}{undetected direct path}
\acrodef{UHF}{ultra-high frequency}
\acrodef{ULA}{uniform linear array}
\acrodef{ULP}{user location protocol}
\acrodef{UMP}{uniformly most powerful}
\acrodef{UMPI}{uniformly most powerful invariant}
\acrodef{UT}{user terminal}
\acrodef{UTC}{coordinated universal time}
\acrodef{UTM}{universal transverse Mercator}
\acrodef{UTRA}{UMTS terrestrial radio access}
\acrodef{UAV}{unmanned aerial vehicle}
\acrodef{UUV}{unmanned underwater vehicle}
\acrodef{UWB}{ultrawide-band}
\acrodef{UWBcap}[UWB]{Ultrawide band}
\acrodef{VFIL}{virtual force iterative localization}
\acrodef{VGA}{variable-gain amplifier}
\acrodef{VNA}{vector network analyzer}
\acrodef{WAF}{wall attenuation factor}
\acrodef{WB}{wideband}
\acrodef{WBI}{wideband interference}
\acrodef{WCL}{weighted centroid localization}
\acrodef{WED}{wall extra delay}
\acrodef{WiMAX} {worldwide interoperability for microwave access}
\acrodef{WLAN}{wireless local area network}
\acrodef{WLS}{weighted least squares}
\acrodef{WMAN}{wireless metropolitan area network}
\acrodef{WPAN}{wireless personal area networks}
\acrodef{WRAPI}{wireless research application programming interface}
\acrodef{WSN}{wireless sensor network}
\acrodef{WSR}{wireless sensor radar}
\acrodef{WSS}{wide-sense stationary}
\acrodef{WWB}{Weiss-Weinstein bound}
\acrodef{WWLB}{Weiss-Weinstein lower bound}
\acrodef{ZZB}{Ziv-Zakai bound}
\acrodef{ZZLB}{Ziv-Zakai lower bound}

\begin{document}
\title{\LARGE{Analysis and Optimization of Reconfigurable Intelligent Surfaces\\ Based on $S$-Parameters Multiport Network Theory}}

\author{ \IEEEauthorblockN{Andrea Abrardo\IEEEauthorrefmark{1}, Alberto Toccafondi\IEEEauthorrefmark{1}, Marco Di Renzo\IEEEauthorrefmark{2}} \\
\IEEEauthorblockA {\IEEEauthorrefmark{1} University of Siena, Siena, Italy \\
Email: abrardo@unisi.it, albertot@unisi.it}\\
\IEEEauthorblockA {\IEEEauthorrefmark{2} Universit\'e Paris-Saclay, CNRS, CentraleSup\'elec, Laboratoire des Signaux et Syst\`emes, France\\
Email: marco.di-renzo@universite-paris-saclay.fr}
\vspace{-1cm}
}

\maketitle

\begin{abstract}
In this paper, we consider a reconfigurable intelligent surface (RIS) and model it by using multiport network theory. We first compare the representation of RIS by using $Z$-parameters and $S$-parameters, by proving their equivalence and discussing their distinct features. Then, we develop an algorithm for optimizing the RIS configuration in the presence of electromagnetic mutual coupling. We show that the proposed algorithm based on optimizing the $S$-parameters results in better performance than existing algorithms based on optimizing the $Z$-parameters. This is attributed to the fact that small perturbations of the step size of the proposed algorithm result in larger variations of the $S$-parameters, hence increasing the convergence speed of the algorithm.
\end{abstract}


\section{Introduction}
Reconfigurable intelligent surface (RIS) is an emerging technology for improving the communication performance of future wireless networks \cite{RenzoZDAYRT20}. For optimization purposes, the RIS elements are often modeled as ideal scatterers, which are capable of introducing any phase shift to the impinging signal without any attenuation. These models are not always electromagnetically consistent, since they do not take into account several aspects that can play an important role in characterizing the operation of a realistic RIS, such as the mutual coupling between the RIS elements and the dependency between the phase and the amplitude of the reflection coefficient \cite{RenzoDT22}, \cite{direnzo2022digital}.

In \cite{DR1}, the authors introduce an operational and end-to-end communication model for the analysis and optimization of RIS-aided wireless systems based on multiport network theory. The model in \cite{DR1} is formulated in terms of $Z$-parameters. Among other factors, the main feature that distinguishes the model in \cite{DR1} from conventional models used for RIS-aided communication systems is the non-linearity, as a function of the tunable impedances, of the end-to-end system response due to the presence of mutual coupling between closely spaced RIS elements. The model in \cite{DR1} has been subsequently used for optimizing the tunable impedances of the RIS, in order to maximize the end-to-end received power \cite{DR2} and the sum-rate in multi-user networks \cite{ABR}. In these works, the non-linearity arising from the mutual coupling is tackled by using the Neumann series approximation, which provides a linearization of the end-to-end transfer function. In a recent work \cite{DR3}, an element-wise iterative approach is proposed, wherein the tunable load impedances are optimized one by one, thus avoding the Neumann approximation that forces the algorithm to proceed using small variations.

An alternative representation for RIS-aided channels based on multiport network theory can be given in terms of $S$-parameters \cite{BC2}, \cite{BC1}. In these papers, however,  the $S$-parameters representation is not further utilized for RIS optimization. In the present paper, motivated by these considerations, we propose an optimization framework to maximize the received power in an RIS-aided single-input single-output (SISO) communication system by using the $S$-parameters representation. In addition, we offer a thorough  comparison of the representation of RIS-aided channels in terms of $Z$-parameters and $S$-parameters. The proposed analysis shows that optimizing an RIS based on the $S$-parameters provides a faster convergence rate, since small variations of the impedance results in larger variations of the corresponding scattering parameters.

\section{System and Channel Model}\label{System model}
We consider an RIS-aided communication channel that consists of a transmitter with $N_T$ antennas, a receiver with $N_R$ antennas, and an $N_S$-port loaded scatterer that models the RIS. The RIS-aided communication link between the transmitter and the receiver can be represented by a linear network with $N=N_T+N_S+N_R$ ports. The $N$-port network can be characterized by a scattering matrix $\mathbf{S}$ that relates the vector $\mathbf{a}$ of the incident power waves on the ports with the vector $\mathbf{b}$ of reflected power waves from the ports, as $\mathbf{b}=\mathbf{S}\mathbf{a}$.

From \cite{BC2}, the $\mathbf{S}$ matrix can be decomposed as
\begin{equation}
    \left[\begin{array}{c}\mathbf{b}_T \\\mathbf{b}_S \\\mathbf{b}_R\end{array}\right]=
    \left[\begin{array}{ccc}\mathbf{S}_{TT} & \mathbf{S}_{TS}& \mathbf{S}_{TR}\\
    \mathbf{S}_{ST} & \mathbf{S}_{SS}& \mathbf{S}_{SR}\\
    \mathbf{S}_{RT}& \mathbf{S}_{RS}& \mathbf{S}_{RR}
    \end{array}\right]
    \left[\begin{array}{c}\mathbf{a}_T \\\mathbf{a}_S \\\mathbf{a}_R\end{array}\right]
    \label{eq:absub}
\end{equation}
where $\mathbf{a}_x$ and $\mathbf{b}_x$ for $x \in [T,S,R]$ are the incident and reflected power wave vectors at the transmitter ($T$), RIS ($S$) and receiver ($R$) ports, respectively. Accordingly,  $\mathbf{S}_{xy}$ for $x,y \in [T,S,R]$ are the scattering sub-matrices that relate the vectors $\mathbf{b}_x$ of the reflected power waves with the vector $\mathbf{a}_y$ of the incident power waves.

In a general configuration, the $N_T$ ports of the transmitter are connected to voltage generators that are characterized by a set of power waves $\mathbf{a}_g=\{a_{g,1},\ldots,a_{g,N_T} \}$ and a set of internal impedances $\mathbf{Z}_g=\{Z_{g,1},\ldots,Z_{g,N_T}\}$. Also, the $N_R$ and $N_S$ ports of the receiver and the RIS are connected to the sets of loads $\mathbf{Z}_R=\{Z_{R,1},\ldots,Z_{R,N_R}\}$ and $\mathbf{Z}_S=\{Z_{S,1},\ldots,Z_{S,N_S}\}$, respectively. Therefore, the sub-vectors in \eqref{eq:absub} are related to the loads by the following expressions:
\begin{equation} \label{eq:atasar}
\mathbf{a}_T= \mathbf{a}_g+\mathbf{\Gamma}_T\mathbf{b}_T; \
\mathbf{a}_S= \mathbf{\Gamma}_S\mathbf{b}_S; \ \mathbf{a}_R= \mathbf{\Gamma}_R\mathbf{b}_R
\end{equation}

Specifically, $\mathbf{\Gamma}_x$ for $x \in [T,S,R]$ is a diagonal matrix with entries $\Gamma_{x,i}=(Z_{x,i}-Z_{0,i})/(Z_{x,i}+Z_{0,i})$, where $Z_{0,i}$ is the reference impedance at the $i$-th port. Usually, $Z_{0,i}=Z_0=\ 50\,\Omega$. To ensure the best power matching at the transmitter and receiver, we consider $\mathbf{\Gamma}_T=\mathbf{\Gamma}_R=0$. Under these conditions, from \eqref{eq:atasar} and \eqref{eq:absub} it is possible to derive the end-to-end channel matrix that relates the incident power wave vector $\mathbf{a}_T=\mathbf{a}_g$ and the output power wave vector $\mathbf{b}_R$, i.e., $\mathbf{b}_R=\mathbf{\hat{H}}_{e2e}\mathbf{a}_T$,  as
\begin{equation}\label{eq:TF}\mathbf{\hat{H}}_{e2e}=\mathbf{S}_{RT} + \mathbf{S}_{RS} \mathbf{\Gamma}_{S}(\mathbf{U}-\mathbf{S}_{SS}\mathbf{\Gamma}_{S})^{-1}\mathbf{S}_{ST}
 \end{equation}
where $\mathbf{U}$ denotes the identity matrix.
By adjusting the matrix $\mathbf{\Gamma}_S$ of reflection coefficients of the RIS, the received power can be optimized.

In \cite[Corollary 1]{DR1}, an end-to-end communication channel model based on the $Z$-parameters is introduced. Since the $S$-parameters and $Z$-parameters matrices are related by the expression \cite{pozar2011} $\mathbf{S}= (\mathbf{Z}-{Z}_{0}\mathbf{U})(\mathbf{Z}+{Z}_{0}\mathbf{U})^{-1}$, the two channel models are equivalent. However, we show next that the $S$-parameter representation may be more convenient when optimizing the end-to-end channel response as a function of the load impedances of the RIS. First, we discuss some interesting features of the $S$-parameters and $Z$-parameters representations for RIS-aided channels based on multiport network theory.

\section{$S$-parameters vs. $Z$-parameters Representations}
The end-to-end channel matrix in \eqref{eq:TF} consists of the sum of two terms, where only the second depends on the loads connected to the ports of the RIS through the diagonal matrix of reflection coefficients $\mathbf{\Gamma}_{S}$. It is worth noting that the first term $\mathbf{S}_{RT}$ depends on the direct transmitter-receiver link and the transmitter-RIS-receiver link when the ports of the RIS are connected to matched loads, which results in $\mathbf{\Gamma}_{S}=0$. Due to the physical presence of the RIS, this latter contribution is always present even if the direct transmitter-receiver link is blocked by the presence of physical obstacles, and it constitutes the structural scattering of the RIS. This is different from the $Z$-parameters representation of the RIS in  \cite[Corllary 1]{DR1}, in which the structural scattering of the RIS is decoupled from the direct link and is embodied into the transmitter-RIS-receiver link. Therefore, the $S$-parameters and $Z$-parameters representations are equivalent, but the terms (and their physical meaning) in the two equations are not one-to-one related.

To better understand this subtle difference, let us consider the  SISO setting, and let us assume that the ports of the RIS are all terminated to the reference impedance $Z_0$ (perfect port matching). From \cite[Corollary 1]{DR1}, we obtain the following (always assuming the best power matching at the transmitter and
receiver):
\begin{equation}\label{eq:TFZ}
 	\mathbf{\hat{H}}_{e2e}(Z_0) = \mathcal{Y}_0\left[Z_{RT}-\mathbf{Z}_{ RS } (\mathbf{Z}_{SS}+{Z}_{0}\mathbf{U})^{-1}\mathbf{Z}_{ ST }\right]
 \end{equation}
where $\mathbf{Z}_{SS}$ is the matrix of self and mutual impendances of the RIS.

If the ports of the RIS are terminated to $Z_0$, we have, as mentioned, $\mathbf{\Gamma}_{S}=\bf{0}$. Thus, the second addend in \eqref{eq:TF} is zero, and, by comparing \eqref{eq:TF} with \eqref{eq:TFZ}, we obtain $S_{ RT } = \mathbf{\hat{H}}_{e2e}(Z_0)$. It is apparent that $S_{ RT }$ depends on the RIS in contrast to $Z_{ RT }$ in \eqref{eq:TFZ}.

To get further engineering insights, let us analyze the case study with no mutual coupling between the RIS elements. Also, let us assume $\mathbf{Z}_{SS}={Z}_{0}\mathbf{U}$ in \eqref{eq:TFZ} but the ports of the RIS are terminated to generic impedances. From $\mathbf{S}= (\mathbf{Z}-{Z}_{0}\mathbf{U})(\mathbf{Z}+{Z}_{0}\mathbf{U})^{-1}$, therefore, we obtain $\mathbf{S}_{SS}=\bf{0}$. Then, \eqref{eq:TF} reduces to $\mathbf{\hat{H}}_{e2e}=\mathbf{S}_{RT} + \mathbf{S}_{RS} \mathbf{\Gamma}_{S}\mathbf{S}_{ST}$. In this case, we retrieve the transfer function that is usually utilized in wireless communications with $\mathbf{\Gamma}_{S} $ denoting the diagonal matrix of reflection coefficients.

Let us analyze again the case study with no mutual coupling and $\mathbf{Z}_{SS}={Z}_{0}\mathbf{U}$, but the ports of the RIS are terminated to $Z_0$. In this case, $\mathbf{\Gamma}_{S}=\bf{0}$ in \eqref{eq:TF}. If $Z_{RT}=0$, from \eqref{eq:TFZ} we obtain
\begin{equation}\label{eq:srtz0}
    S_{RT} = \mathbf{\hat{H}}_{e2e}(Z_0) = - \frac{\mathcal{Y}_0}{2Z_0}\mathbf{Z}_{ RS } \mathbf{Z}_{ ST }
\end{equation}

Therefore, we see that, even if the direct transmitter-receiver link is blocked by physical objects and the reflection coefficient of the RIS is equal to zero, the received signal is not equal to zero, since $S_{RT} \ne 0$. This terms is equivalent to a specular reflection from the RIS and is the so-called structural scattering, which occurs when all the ports of the RIS are terminated to $Z_0$ \cite{knott2004}. In communication papers, this term is usually ignored: In fact, if the direct transmit-receiver link is blocked and the reflection coefficient is zero, it is assumed that the received signal is zero as well. It is worth noting that in the $S$-parameters model in \eqref{eq:TF}, the structural scattering is contained in $S_{RT} \ne 0$. In the $Z$-parameters model in \eqref{eq:TFZ}, on the other hand, the structural scattering is not contained in $Z_{RT}$, but in the RIS-dependent term.

Finally, let us consider \eqref{eq:TF} and let us assume that $\mathbf{S}_{ SS }$ is a diagonal matrix for simplicity. Even if the reflection coefficients in $\mathbf{\Gamma}_S$ are assumed to have a unit modulus, we evince that the presence of the RIS has an impact on the amplitude and phase of the incident signal. Also, the impact of the phase and the amplitude cannot be decoupled. This is due to the term $\mathbf{\Gamma}_{S}(\mathbf{U}-\mathbf{S}_{SS}\mathbf{\Gamma}_{S})^{-1}$. If the elements of $\mathbf{\Gamma}_S$ have a unit modulus, the RIS operates only on the phase of the incident signal if and only if $\mathbf{S}_{ SS } = \bf{0}$. If $\mathbf{S}_{ SS }$ is not a diagonal matrix, the relationship between the amplitude and the phase is, in general, even more pronounced.

 \begin{figure*}\hspace{-1cm}
\begin{minipage}[t]{1\linewidth}
\centering
\includegraphics[scale=.45]{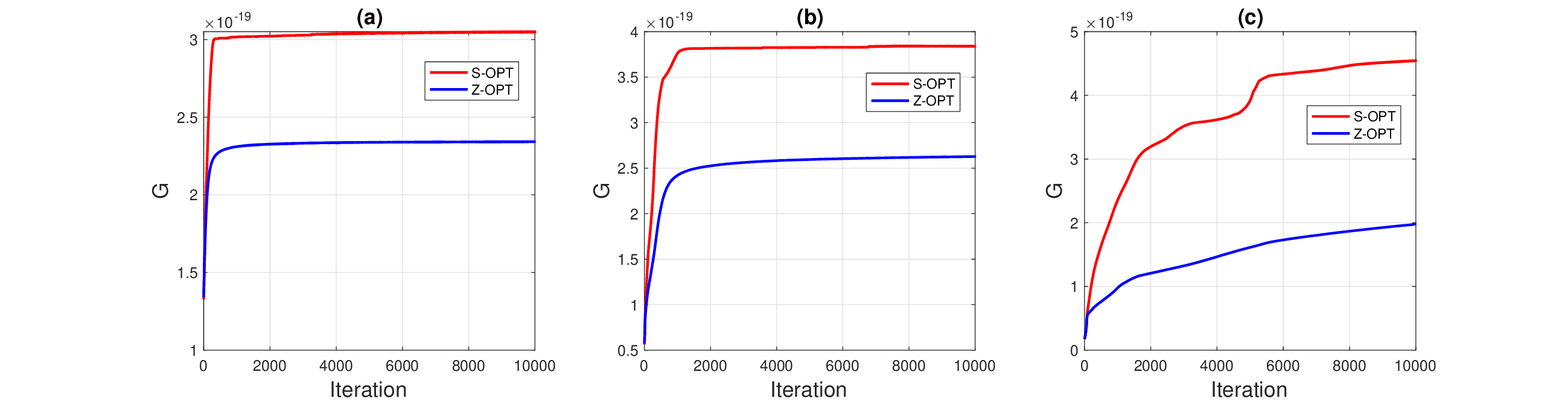}\\
\caption{Convergence vs. iterations: (a) $d_y = \lambda/4$; (b) $d_y = \lambda/8$ ; (c) $d_y = \lambda/16$}
\label{fig1}
\end{minipage}
\vspace{-0.3cm}
\end{figure*}

 \section{RIS optimization}

The RIS optimization problem is formulated for the SISO case, i.e., $N_T = N_R = 1$. In this case, \eqref{eq:TF}
can be written as
 \begin{equation}\label{eqTF2}
 	{b}_R= \left[{S}_{RT} + \mathbf{S}_{RS} \mathbf{\Gamma}_{S}(\mathbf{U}-\mathbf{S}_{SS}\mathbf{\Gamma}_{S})^{-1}\mathbf{S}_{ST} \right]{a}_T
 \end{equation}

In this setting, we consider the optimization problem
\begin{align}\label{Optprob1}
&\max \limits_{\mathbf{\Gamma}}  \left|{S}_{RT} + \mathbf{S}_{RS} \mathbf{\Gamma}(\mathbf{U}-\mathbf{S}_{SS}\mathbf{\Gamma})^{-1}\mathbf{S}_{ST} \right|^2  \\
&\text{s.t.} \quad \left|\Gamma_{k}\right| = 1 \text{ for } k=1,\ldots,N_S \nonumber
\end{align}
where $\mathbf{\Gamma} = \mathbf{\Gamma}_{S}$ and $\Gamma_{k} = \Gamma_{k,k}$ to simplify the notation.

The main challenge in solving \eqref{Optprob1} is that the objective function depends on the inverse of a matrix, which in turn depends on $\mathbf{\Gamma}$ that is the optimization variable. This makes the problem strongly non-convex. To circumvent this problem, we use an iterative algorithm that provably allows us to increase the objective function at each iteration and hence to find a local optimum for the problem at hand.

To elaborate, we denote by $X_k$ the $k$-th tunable reactance of the RIS and by $\Sigma_{k} = \frac{j{X}_k-1}{j{X}_k+1} = e^{j\phi_k}$. Also, we assume that the parasitic resistances of the tunable impedances of the RIS are much smaller than $Z_0 = 50 \, \Omega$.

The algorithm works iteratively and adjusts $\phi_k$ at each iteration, by introducing small perturbations with respect to the previous iteration, i.e., $\phi_k^{(m+1)} = \phi_k^{(m)} + \delta_k^{(m)}$. Under these assumptions, the matrix inversion can be linearized by using the Neumann series approximation, and the problem in \eqref{Optprob1} can be solved. Specifically, with the aid of some mathematical manipulations, one can show that the optimum $\delta_k^{(m)}$ at the $m$-th iteration is
 \begin{align}\label{solution1}
&\delta_k^{(m)} = \left\{\begin{array}{cc}
      \Delta_m &  \text{if }  \left| A^{(m)} +  C_k^{(m)} \Delta_m \right| \ge \left| A^{(m)} -  C_k^{(m)} \Delta_m \right| \\
      -\Delta_m & \text{otherwise}
\end{array}\right.
\end{align}
where $\Delta_m \ll 1$ is the step-size, $\mathbf{\Gamma}^{(m)}$ is the vector of reflection coefficients at the $m$-th iteration, $A^{(m)} = {S}_{RT} + \mathbf{S}_{RS}\mathbf{Q}_m^{-1}\mathbf{S}_{ST}$, $\mathbf{Q}_m = \left(\mathbf{\Gamma}^{(m)}\right)^{-1} -\mathbf{S}_{SS}$, $C_k^{(m)} = b_{1,k}^{(m)}b_{2,k}^{(m)}$, where $b_{1,k}^{(m)}$, $b_{2,k}^{(m)}$ are the entries of $\mathbf{b}_1^{(m)} = \mathbf{S}_{ RS }\mathbf{P}_m$ and $\mathbf{b}_2^{(m)} = \mathbf{Q}_m^{-1} \mathbf{S}_{ ST }$, with $\mathbf{P}_m = -j \mathbf{Q}_m^{-1}\left(\mathbf{\Gamma}^{(m)}\right)^{-2}e^{j\boldsymbol{\phi}^{(m)}}$, where $e^{j\boldsymbol{\phi}^{(m)}}$ is a diagonal matrix whose entries are $e^{j{\phi}_k^{(m)}}$.

\section{Numerical Results}\label{sec:4}
We consider a carrier frequency $f = 28$ GHz, the RIS is centered at the position $(0,0,2)$ m, and the transmitter is located at the position $(4,0,3)$ m. The receiver is located at the position $(0,7, 4, 1)$ m, i.e., it is located at an angle of nearly 80 degrees with respect to the direction of specular reflection. The transmitting and receiving antennas, as well as the scatterers of the RIS are identical perfectly conducting $z$-directed dipoles with radius $\lambda/500$ and length $L = 0.46 \lambda$. The RIS dipoles are arranged as a uniform planar array, and the spacing of the elements is $d_y = \lambda/Q$ with $Q = 4,8,16$ in the $y$ direction and $d_z = 3/4 \lambda$ in the $z$ direction. The RIS has approximately the shape of a square with an area $4 \lambda^2$.

In Fig. \ref{fig1}, we report the convergence of the proposed algorithm, which is denoted by S-OPT, and compare it against the benchmark scheme proposed in \cite{DR2}, which operates on the $Z$-parameters representation and is denoted by Z-OPT. The comparison is given in terms of the squared norm of the end-to-end transfer function in Sec. \ref{System model}, which is denoted by $G$. As in \cite{DR2}, we assume that the direct link is negligible due to the presence of obstacles between the transmitter and receiver.

It is observed that S-OPT outperforms Z-OPT.
This is attributed to the fact that small perturbations of the step size of S-OPT result in larger variations of the $S$-parameters, hence increasing the convergence speed of the algorithm.

\section{Conclusion}
In this paper, we have compared the $Z$-parameters and $S$-parameters representations of an RIS-aided channel. Also, we have developed an algorithm for optimizing the RIS configuration in the presence of electromagnetic mutual coupling based on the $S$ parameters, which is shown to outperform existing algorithms based on optimizing the $Z$ parameters.

\bibliographystyle{IEEEtran}
\bibliography{Biblio,references_DD,reference}

\end{document}